\documentstyle[12pt]{article}
\begin{document}
\baselineskip=22pt plus 0.2pt minus 0.2pt
\lineskip=22pt plus 0.2pt minus 0.2pt
\begin{center}
 \Large
 On the gauge fixing of 1 Killing field reductions of canonical gravity:
the case of asymptotically flat induced 2-geometry\\

\vspace*{0.35in}

\large

Madhavan Varadarajan
\vspace*{0.25in}

\normalsize

Relativity group, University of Utah, Salt Lake City, UT 84112

\vspace{.5in}
March 2, 1995\\
\vspace{.5in}
ABSTRACT
We consider 1 spacelike Killing vector field reductions of 4-d vacuum
general relativity. We restrict attention to cases in which the manifold of
orbits of the Killing field is $R^{3}$. The reduced Einstein equations are
equivalent to those for Lorentzian 3-d gravity coupled to an SO(2,1) nonlinear
sigma model on this manifold. We examine the theory in terms of a Hamiltonian
formulation obtained via a 2+1 split of the 3-d manifold. We restrict
attention to geometries which are asymptotically flat in a 2-d sense defined
recently. We attempt to pass to a reduced Hamiltonian description in terms
of the
true degrees of freedom of the theory via gauge fixing conditions of 2-d
conformal flatness and maximal slicing. We explicitly solve the
diffeomorphism constraints and relate the Hamiltonian constraint to the
prescribed negative curvature equation in $R^2$ studied by mathematicians.
We partially address issues of
existence and/or uniqueness of solutions to the various elliptic partial
differential equations encountered.

\end{center}

\pagebreak

\setcounter{page}{1}

\section{Introduction}

To develop technical tools for, as well as to address conceptual questions
which arise in , the efforts to build a quantum theory of 4-d gravity, it is
of use to study simple models which retain some of the features of the full
theory.  An interesting model  which captures some of the diffeomorphism
invariance as well as the nonlinear {\em field} theoretic character of
full gravity,
is the midisuperspace of 1 spacelike Killing vector field (henceforth
referred to as  1kvf) reductions of 4-d vacuum general relativity.

We beleive that it is essential to study the classical behaviour of a system
before trying to discover the underlying quantum theory. In this paper we
study some features of the classical theory of 1 kvf reductions of vacuum
general relativity.

It is known that the Einstein equations for 4-d spacetimes with 1kvf are
 equivalent to the equations describing 3-d gravity suitably coupled to
2 scalar fields \cite{exact}. The 3-d manifold is the manifold of orbits of
the kvf and
the scalar fields are related to the norm and twist of the kvf. In the case
in which the 3-d spacetime admits a 2+1 split, one can construct a Hamiltonian
framework for the system. Moncrief has studied the case in which the
2-manifold
in the 2+1 split is compact in \cite{vince}. In this paper we try to
generalize his work to the noncompact 2 manifold case, when the 2-geometry
is asymptotically flat.
We use the notion of asymptotic flatness developed in \cite{asht}. We try to
gauge fix
the theory and develop a reduced Hamiltonian description for the true degrees
of freedom along the lines of \cite{vince}. Apart from the general reasons
outlined
above, there is a more model specific reason for this work which we outline
below.

Recently a Hamiltonian framework for asymptotically flat 2+1 gravity
coupled to smooth matter fields was developed in \cite{asht}. The generator
of
time translations at spatial infinity was identified as the energy of the
system and it was shown (as in the point particle case \cite{jack} which
dealt with
non smooth matter fields) that this energy
was bounded from above. There have been claims \cite{gamba} that the
perturbative
{\em quantum} theory of 2+1 gravity coupled to scalar fields is
renormalizable.
A natural question to ask, is whether the boundedness of the energy has
anything to do with this good ultraviolet behaviour. The treatment
in \cite{gamba} did not
explicitly take into account the upper bound on energy (Also it
 is not clear to us
whether \cite{gamba} dealt with a single scalar field or with an order 1/N
expansion).

Hence, we would like to study the classical theory of 1kvf reductions
 from a 2+1 perspective, in such a way as to bring to the forefront, the
boundedness of the Hamiltonian, so that quantization attempts could deal
more directly with this issue. So much for motivation.

Let us briefly summarize the results of this paper. We use and slightly extend
the Hamiltonian
framework developed in \cite{asht}. Viewing our midisuperpace as 2+1 gravity
coupled to matter, we impose the gauge fixing conditions of 2-d conformal
flatness (note that we deal only with the case of the 2-manifold being $R^2$)
 and maximal slicing. This allows us to solve the diffeomorphism constraints
and the Hamiltonian constraint becomes an elliptic partial differential
equation  for the conformal factor, much as in \cite{vince}.
{}From \cite{asht}, it is apparent that the asymptotic behaviour of the
conformal factor determines the true Hamiltonian of the system. The partial
differential equation
for the conformal factor has been studied by mathematicians (see for eg.
\cite{mcowen}) and we quote
some of their results on existence of solutions. These results have
intriguing connections with the upperboundedness of the Hamiltonian. The
final picture is of the matter fields (which describe the true degrees of
freedom) being evolved in time by the true time independent
Hamiltonian of the system.

The one unattractive result of this analysis is that the lapse and
shift fields diverge as spatial infinity is approached. One may question
as to whether such foliations are acceptable from a spacetime point of view.
However, it should be noted that the Hamiltonian framework {\em does admit}
evolutions generated by  such lapses and shifts. We shall say more about
this later.

The layout of this paper is as follows. Section 2 is devoted to a slight
extension of the Hamiltonian framework of \cite{asht} and the assertions
made above,
about the lapse and shift are proved here. In section 3, we describe the
equations governing the midisuperspace and display the gauge fixing
conditions.
In section 4 we deal with the diffeomorphism constraint and in section 5 with
the Hamiltonian constraint. Section 6 deals with the propagation of the
gauge conditions and the associated elliptic pdes for the lapse and shift.
Section 7 deals with
 the reduced Hamiltonian description and also describes a  vague idea
related to perturbative
quantization.
Section 8 contains conclusions and open questions.

We shall set $k=c=1$ in this paper, $k$ being the gravitational constant in
2+1 dimensions.

\section{The Hamiltonian framework for asymptotically flat 2+1 gravity}

We use the same notation as in \cite{asht}.
Although the main part of the paper deals with the case of
the spatial 2 manifold $\Sigma$ being $R^2$, for this section we shall only
require that $\Sigma$ be noncompact with ($\Sigma$- a compact set)
diffeomorphic
to ($R^2$ - a compact set). We shall use slightly more general boundary
conditions than
in \cite{asht}. Due to the similarity of analysis here and in  \cite{asht},
we shall be brief and
only highlight the new results obtained here as a result of choosing more
general
boundary conditions.

The 3-d spacetime has topology $\Sigma \times R$; the phase space variables
are
the 2-metric $q_{ab}$ and its conjugate momentum $P^{ab}$. We fix a flat
metric
$e_{ab}$ in the asymptotic region of $\Sigma$ diffeomorphic
to $R^2$. $(r,\theta)$
are its polar coordinates $(0 \leq \theta \leq 2\pi)$
and spatial infinity is approached as
$r\rightarrow \infty$.
The asymptotic behaviour of $q_{ab}$ is
\begin{equation}
q_{ab}= r^{-\beta}[e_{ab}  + O(1/r^{\epsilon})] \;\;\;\; \epsilon >0
\end{equation}
In \cite{asht} we had fixed $\epsilon=1$, but here we allow $\epsilon$ to be
arbitrarily small. Since $P^{ab}$ coordinatizes cotangent vectors to the
space
of 2-metrics $q_{ab}$,
\begin{equation}
P[\delta q] =\int_{\Sigma} d^2 x P^{ab} \delta q_{ab}
\end{equation}
with
\begin{equation}
\delta q_{ab} \sim -\delta \beta (\ln r)
             r^{-\beta}[e_{ab}  + O(1/r^{\epsilon})]
              \; + \; r^{-\beta} O(1/r^{\epsilon}) ,
\end{equation}
should be well defined.
 This fixes the behaviour of $P^{ab}$ near spatial infinity to be
\begin{equation}
P^{ab}e_{ab} \sim r^{\beta-2- \delta} \;\;\;\;
[P^{ab}-1/2 Pq^{ab}] \sim r^{\beta-2}
\end{equation}
where $\delta>0$ and can be arbitrarily small.
(As in \cite{asht},$P=P^{ab}q_{ab}$.)

A few useful fall offs induced by those already mentioned above are
\begin{equation}
\sqrt{q} \sim r^{-\beta} \;\;\;\sqrt{q} R \sim r^{-(2+\epsilon)},
      \epsilon >0
\end{equation}
where $R$ is the scalar curvature of $q_{ab}$.

Note that we assume , as in \cite{asht}, that our matter fields are of
compact support.

\subsection{The diffeomorphism constraints}

Given a shift $N^a$ on $\Sigma$, the smeared diffeomorphism  constraints
can be written as:
\begin{equation}
C_{\vec{N}}  = -2\int_{\Sigma} d^2x\  N^{a}D_{c}(P^{cd}q_{da})
+ \;\;\;{\rm{matter \; \; terms}}
\end{equation}
With our assumptions on the matter fields, the integral involving matter
fields is well-defined and will play no role in the discussion of this
section.
We will therefore focus just on the gravitational part, i.e., the first term
on the right hand side of (6), which we will refer to as
$C_{\vec N}^{\rm geo}$.
It can be verified, following an analysis similar to
that in \cite{asht}, that for $N^{a} \sim r^{1-\alpha}, \;\;\; \alpha >0$:\\
\noindent (a)$C_{\vec N}^{\rm geo}$ is well defined \\
\noindent (b)$C_{\vec N}^{\rm geo}$ is functionally differentiable
on the phase space. \\
\noindent (c)$C_{\vec N}^{\rm geo}$ generates infinitesmal diffeomorphisms
on
$(q_{ab} , P^{ab})$
which do not take $(q_{ab} , P^{ab})$  out of the phase space i.e.
the new $(q_{ab} , P^{ab})$  also respect the asymptotic conditions.

Hence, from a constraint theory view point, these `exploding' diffeomorphisms
must be considered as gauge, since they are generated by first class
constraints. Thus, the picture is in sharp contrast to
3+1 dimensions where the diffeomorphisms considered as gauge are all trivial
at infinity. We see here, that what seems reasonable from a constraint theory
viewpoint may not be so from a spacetime viewpoint.

\subsection{ The Hamiltonian constraint}

Given a lapse function $N$ on $\Sigma$, we can write the smeared
constraint function as:
\begin{eqnarray}
C_N & =  & - \int_{\Sigma} d^2x  N [\sqrt{q}R -
{1\over\sqrt{q}}(P^{ab}P_{ab}
   - P^{2})]+ {\rm matter \;\; terms} \\
   & =  & C_{N}^{\rm geo}  + C_{N}^{\rm matter}
\end{eqnarray}

Again, matter terms will play no role in our discussion.
We note the following with regard to existence of $C_{N}^{\rm geo} $ :\\
\noindent (a)  For ($N \rightarrow$ constant) near spatial infinity,
the integral is well defined  only when $\beta < 2$. For $\beta >2$,
as in \cite{asht}, the contribution due to the kinetic terms diverges; the
potential
term (see equation(5)) always falls off faster than $1/r^2$ and poses no
problem. \\
\noindent (b) For $\beta<2$  {\em even if } $N \rightarrow \ln r$ as
$r \rightarrow \infty$, $C_{N}^{\rm geo}$ is well defined!

Let us now turn to differentiability of $C_{N}^{\rm geo} $. The kinetic
terms pose no problem. We examine only the potential term. It can be
checked that
\begin{eqnarray}
\delta \int_{\Sigma} d^2x N \sqrt{q} R &  =  \int_{\Sigma}d^2x
                   \sqrt{q}(-D_{a}D_{b}N
                           + D_{c}D^{c}Nq_{ab}) \delta q^{ab} \nonumber \\
                              \;\;\;  & +\oint_{r=\infty} d\theta
                            \sqrt{c} [Nv_{a}+(D_{a}N)q^{bd}\delta
               q_{bd}-D^{c}N\delta q_{ac}] r^{a},
\end{eqnarray}
where,
\begin{equation}
 v_{a}=D^{b}\delta q_{ab}-D_{a}(q^{bd}\delta q_{bd}),
\end{equation}
$r^a$ is the unit normal to the circle at spatial infinity and $\sqrt{c}$
is the determinant of the induced metric, $c_{ab}$, on this circle.
The integrals are to be understood in the sense of \cite{asht}.

Now consider the following asymptotic behaviour for $N$.\\
\noindent (1) If $N\sim 1/r^{\eta}, \;\;\; \eta >0$,
it is easy to check that the surface term vanishes.\\
\noindent (2) If $N\sim N_{\infty} + O(1/r^{\eta}), \;\;\; \eta >0$,
$N_{\infty}$ being a constant, as in \cite{asht} the surface term does not
vanish
and we need to add a term $\delta \beta 2\pi N_{\infty} $ to the constraint
functional to make the resulting functional differentiable on phase space.
 This leads, exactly as in \cite{asht}, to the identification of $\beta /8$
with the total energy of the system. \\
\noindent (3) If $N\sim  \ln r \;+\; O(1/r^{\eta}), \;\;\; \eta >0$,
it is straightforward to verify that the surface term {\em vanishes}!
This is due
to the fact that both $N$ and $q^{ac} \delta q_{ab}$ diverge as $\ln r$
near spatial infinity.
Again, this is in sharp contrast to what happens in 3+1 dimensions. Once again
 a strictly constraint systems viewpoint would identify such motions as gauge,
even though it seems strange from a spacetime viewpoint.

The evolution generated by the constraint functionals on $q_{ab}, P^{ab}$
is given below:
\begin{eqnarray}
\dot{q}_{ab} &  = & 2N{1\over \sqrt{q}} (P_{ab} - Pq_{ab})
                      + {\cal {L}}_{\vec{N}} q_{ab}      \\
\dot{P}^{ab} & =  & \sqrt{q} [D^{a}D^{b}N- q^{ab}D^{c}D_{c}N]
                      + {q^{ab}\over \sqrt{q}}N[P^{ef}P_{ef}-P^{2}]
\nonumber\\
\;\;         & \; & -{2N\over \sqrt{q}} [P^{ac}P_{c}^{b}-P P^{ab}]
                      +  {\cal {L}}_{\vec{N}} P^{ab} +
{\rm matter \; terms}
\end{eqnarray}

Note that for infintesimal evolutions generated by the (first class)
Hamiltonian constraint smeared with
\begin{equation}
N\sim  \ln r \;+\;
O(1/r^{\eta}), \;\;\; \eta >0,\;\;r\rightarrow\infty ,
\end{equation}
the phase space variables
$(q_{ab}(t), P^{ab}(t))$ do respect the asymptotic conditions and
and these infinitesmal evolutions must
be identified as gauge from a constraint system  viewpoint.

For a lapse with asymptotic behaviour $N\rightarrow a\ln r \;+ b\;
                + \;O(1/r^{\eta})$ , $a,b$ being constants, the
above evolution equations are generated by the true Hamiltonian functional
\begin{equation}
H= 2\pi b \beta + \int_{\Sigma}NC d^2 x \approx 2\pi b \beta
\end{equation}
and again, infinitesmal evolutions preserve the boundary conditions. These
evolutions are {\em not} identified with gauge. Note that in \cite{asht},
there
was an inconsistency, in that the evolution generated by the true
Hamiltonian did not necessarily preserve the boundary conditions on $q_{ab}$.
 By not fixing
$\epsilon = 1$ in equation (1),
we have rectified this inconsistency and also shown that the
upper bound on the energy still exists.

\section{The action and constraints for the 1kvf midisuperspace}

The discussion in the remainder of the paper
 could, in principle apply to arbitrary matter couplings
of compact support, but we restrict ourselves to those which come from the
1kvf reduction of 4-d general relativity (see \cite{vince}). We shall, from
 now on, restrict
our considerations to the case where the spatial manifold, $\Sigma$,
is  $R^2$.
 The action $S$ and the constraints $C_{a}$ and $C$ are
\begin{equation}
S= \int dt[ (\int_{R^2}  d^2 x [ P^{ab}{\dot {q}}_{ab} +s \dot{\gamma}
            +v \dot{\omega} -NC -N^{a}C_{a}]) - 2\pi\beta]
\end{equation}

\begin{eqnarray}
C_{a} & = & -2D_{b}P^{b}_{a} \; +\; s{\gamma}_{,a}\; +\;\;v{\omega}_{,a}\\
C     & = & \sqrt{q} [-R\; +\; 2 q^{ab}{\gamma}_{,a}{\gamma}_{,b}
                + {1 \over 2}e^{-4\gamma}q^{ab}{\omega}_{,a}{\omega}_{,b}]
                      \nonumber \\
      & + & {1\over \sqrt{q}}[P_{ab}P^{ab}- P^{2} + {1\over 8}s^{2}
                   +{1\over 2}e^{4\gamma} v^{2}]
\end{eqnarray}
 Here $\gamma, \omega$ are the scalar fields obtained as a result of the
kvf reduction and $s,v$ are their conjugate momenta.

As mentioned earlier, all the matter fields and momenta have compact support.
The asymptotic behaviours of $(q_{ab} , P^{ab})$ as well as the shift $N^{a}$
have been given in section 2.
The lapse
\begin{equation}
N \sim a \ln r + 1 + O(1/r^{\eta}),\;\; \eta >0
\end{equation}
near infinity.

\subsection{The gauge fixing conditions}

We follow Moncrief's \cite{vince} ideas for gauge fixing by adopting the
York procedure \cite{york}.
The system has 2-d diffeomorphism invariance as well as time
reparameterizations
of the type generated by the Hamiltonian constraint. But in contrast
to \cite{vince},
we do have a sense of time at infinity. Thus we do not need to deparameterize
the theory  as in \cite{vince} , but can proceed to gauge fix the system
completely.
We do this by  choosing foliations of the 3-d spacetime by maximal slices i.e.
\begin{equation}
P=0
\end{equation}
and demanding that the dynamical 2- metric be conformal to a fixed flat
metric
$h_{ab}$ (this is reasonable, since the 2-manifold is $R^2$):
\begin{equation}
q_{ab}= e^{2\lambda} h_{ab}
\end{equation}
 As mentioned earlier we do not know the conditions under which 3-d
spacetimes
admit maximal slices. We now examine how the asymptotic conditions interact
with the gauge fixing conditions (19) and (20).

Clearly, $P=0$ is admitted by our boundary conditions. With regard to the
conformal flatness condition we impose
\begin{equation}
h_{ab} \rightarrow e_{ab} \;\;\;\; r\rightarrow \infty
\end{equation}
This implies the asymptotic behaviour
\begin{equation}
\lambda \rightarrow {-\beta \over 2} \ln r \;
+\; O(1/r^{\epsilon}) \;\;\;\epsilon >0
\end{equation}
 Note that there is no nonvanishing constant piece in $\lambda$ near infinity.
This will be important for later considerations.

We fix cartesian coordinates $(x^{1}, x^{2})$ , associated with $h_{ab}$
and demand that they agree with those of $e_{ab}$ near infinity.
This will be important
for considerations in section 6. We shall work with these coordinates in
future calculations.  We shall denote by $i,j..=1,2$ the corresponding
cartesian components of tensors. All densities will be evaluated using
these cartesian
coordinates. Thus, for example, $\sqrt {h}=1$.

We
 denote the derivative operator compatible with $h_{ab}$ by $\partial_{a}$.

All abstract indices a,b.. will be raised and lowered by $q_{ab}$, unless
otherwise
mentioned.

We decompose $P^{ab}$ as in \cite{vince}, in accordance with the York
procedure:
\begin{equation}
P^{ab} = {1\over 2}P q_{ab}\; +\;
             e^{-2\lambda}\sqrt{q}(D^{b}Y^{a}+D^{a}Y^{b}-q^{ab}D_{c}Y^{c})
                            \; -\; B^{ab}
\end{equation}
where $B^{ab}$ denotes the transverse traceless part of $P^{ab}$.

 \vspace{4mm}

\noindent {\bf Claim:} $B^{ab}=0$\\
\noindent {\bf Proof:} Since $B^{ab}$ is a transverse traceless symmetric
tensor density of weight 1:
\begin{eqnarray}
D_{a}B^{a}_{b} & = & \partial_{a}B^{a}_{b}=0\\
\Rightarrow \partial_{a}(B^{a}_{b}X^{b}_{i}) & = & 0
\end{eqnarray}
where  $X^{b}_{i}=({\partial \over {\partial x^{i}}})^{b}$ is
the translational Killing vector field of $h_{ab}$ in the $i^{th}$ direction.
Define
\begin{equation}
B^{a}_{i}:=B^{a}_{b}X^{b}_{i} \;\;\;\;\; w_{ib}:=n_{ab}B^{a}_{i}
\end{equation}
     where $n_{ab}$ is the Levi Civita antisymmetric tensor density of
weight -1.
\begin{equation}
\Rightarrow \partial_{a} w_{ib}- \partial_{b} w_{ia}=0
\Rightarrow w_{ia}=\partial_{a} w_{i}
\end{equation}
where $w_i$ are scalar functions and we have used equation (25) and
the fact that the 2-manifold
is $R^2$. Using the tracefree and symmetric properties of $B^{ab}$ we get
\begin{equation}
\partial_{1}w_{2}= \partial_{2} w_{1}\;\;,\;\partial_{2} w_{2}=
                                              -\partial_{1}w_{1}
\end{equation}
\begin{equation}
\Rightarrow \Delta w_{i}=0
\end{equation}
where $\Delta$ is the flat space Laplacian operator. From the
boundary conditions
on $P^{ab}$,
\begin{equation}
B^{a}_{b}\sim 1/r^{2} \Rightarrow \partial_{i} w_{j} \sim 1/r
               \Rightarrow  w_{j}\sim c_{j}
\end{equation}
where $c_{j}$ are constants.
The associated Dirichlet problem for equation (29) has a unique solution
$w_{i}=c_{i}$ and hence $B^{a}_{b}=0$

\section{The diffeomorphism constraints}
 The diffeomorphism constraints are of the form
\begin{equation}
D_{a}P^{a}_{b} = M_{b},\;\; {\rm where}\;\;  M_{b}= {1\over 2}
                (s\partial_{b}\gamma+v\partial_{b}\omega)
\end{equation}
 There is an integrability condition for this equation as shown below.

 Note that since $P^{a}_{b}$ is a symmetric traceless tensor density of
weight
1,
\begin{equation}
D_{a}P^{a}_{b}= \partial_{a}P^{a}_{b}
\end{equation}
Using the same notation as in the proof of the claim above, we have
\begin{equation}
\int_{R^2}(\partial_{a} P^{a}_{b}) X^{b}_{i}d^2 x =
   \oint_{r=\infty}P^{a}_{b} X^{b}_{i} {\stackrel{\circ}{r}}_{a} r d \theta
         -\int_{R^2}(\partial_{a}X^{b}_{i}) P^{a}_{b}d^2 x
\end{equation}
where ${\stackrel{\circ}{r}}_{a}$ is the unit radial normal in the flat metric
$h_{ab}$ to the circle at spatial infinity.
The second term on the right hand side above,
vanishes since $X^{b}_{i}$ are Killing vectors of $h_{ab}$.
The first term on the right hand side vanishes because $P^{i}_{j}\sim 1/r^2$
near infinity. Thus, the integrability conditions for the diffeomorphism
constraints are
\begin{equation}
\int_{R^2}M_{i} d^2 x =0
\end{equation}
 Using the gauge conditions  and $B^{ab}=0$, we substitute the York
decomposition (23), into the diffeomorphism constraints and obtain
\begin{equation}
\Delta (Y^{j} h_{ij})= M_{i}
\end{equation}
Note that $P^{a}_{b}\;$ is unchanged if one adds a conformal kvf of $q_{ab}$
(or equivalently, of $h_{ab}$ )
to $Y^{a}$. Since $P^{a}_{b} \sim 1/r^2$ near infinity and we are not
concerned about ambiguities in $Y^{a}$ stemming from addition of conformal
kvfs of
$h_{ab}$, we look for solutions to (35) such that $Y^{i}\rightarrow 0$ as
$r\rightarrow \infty$. Again, the associated Dirichlet problem has a
unique solution
\begin{equation}
Y^{j}h_{ij}= {1\over 2\pi}\int_{R^2} M_{i}(y) \ln |\vec{x}-\vec{y}| d^2 y
\end{equation}
where the integrability condition (34), ensures that $Y^{j}h_{ij}\sim 1/r$
near infinity and we define
$ (|\vec{x}-\vec{y}|)^2 := h_{ij}(x^{i}-y^{i})(x^{j}-y^{j})$.

Thus $P^{i}_{j}$ is uniquely given by
\begin{eqnarray}
P^{i}_{j} & = & {1\over 2\pi}\int_{R^2} {d^2 y \over |\vec{x}-\vec{y}|^2}
               [M_{j}(y)(x^{i}-y^{i})
                       + M_{k}(y)(x^{l}-y^{l})h_{lj}h^{ki} \nonumber \\
          &   &    - \delta^{i}_{j} M_{k}(y)(x^{k}-y^{k})]
\end{eqnarray}
Here $h_{ij} =\delta_{ij}$ (the Kronecker delta function) and $h^{ij}$
denotes the inverse to $h_{ij}$. It can be checked that the above expression
for  $P^{i}_{j}$
satisfies the relevant asymptotic behaviour by virtue of the integrability
 condition (34).  Thus, we have solved the diffeomorphism constraints.

\section{The Hamiltonian constraint}

The Hamiltonian constraint becomes an elliptic nonlinear partial differential
equation for $\lambda$
\begin{equation}
\Delta \lambda  + 2\pi g(x) + f(x) e^{-2\lambda} =0
\end{equation}
with  (note that $h^{ab}$ below , denotes the flat contravariant metric)
\begin{eqnarray}
2 \pi g(x) & = & 2 h^{ab} (\gamma_{,a} \gamma_{,b} \;+ \;
        {e^{-4\gamma}\over 2} \omega_{,a} \omega_{,b}),\;\; \;\; g(x)\geq 0\\
f(x) & = & P^{a}_{b} P^{b}_{a} +{1\over 8}s^2 +  {e^{4\gamma}\over 2}v^2
               \;\;\;\; f(x)\geq 0
\end{eqnarray}
We choose to analyze the above equation for the two exhaustive cases of
$f(x)=0$ and $f(x)$ not identically zero.

\subsection{$f(x)$=0}
This happens if and only if the matter momenta vanish i.e.
$s=v=0$.  The equation then reduces to
\begin{equation}
\Delta \lambda = -2 \pi g(x)
\end{equation}
For $\lambda$ satisfying (22), a solution to this equation is
\begin{equation}
\lambda_{1} = -\int_{R^2}g(y) \ln |\vec{x}-\vec{y}| d^2 y
\end{equation}
This has asymptotic behaviour
\begin{equation}
\lambda \sim -(\int_{R^2} g(y)d^2 y)\ln r \; +\; O(1/r)
\Rightarrow \beta :=\beta_{1} =2\int_{R^2} g(y)d^2 y
\end{equation}
Note that for large enough scalar field strengths, $\beta > 2$
is possible. Such initial data would not be acceptable.
Further, note that (42) is the unique solution to (41)  satisfying (22).
To see this, let $\lambda_{2} \neq \lambda_{1} $ be a solution to (41)
with, in obvious notation, $\beta _2$ not necessarily equal to $\beta_{1}$.
An application of the
maximum principle \cite{gilbarg}  to the Dirichlet problem associated  with
($\lambda_2 -\lambda_1)$ shows that $\lambda_1 =\lambda_2$ is the
only possible
well behaved solution to (41).

Thus the energy, $\beta$, is completely
determined by the Hamiltonian constraint equation.

\subsection{$f(x)$ not identically vanishing}

We proceed to transform the equation  (38) to a desired form.
Let $\lambda = \lambda_{0} +\lambda_{1}$ where
\begin{equation}
\lambda_{1} = -\int_{R^2}g(y) \ln |\vec{x}-\vec{y}| d^2 y
\;\;\; \Rightarrow  \Delta \lambda_{1}=-2\pi g(x)
\end{equation}
As before
\begin{equation}
\lambda_{1} \sim -(\int_{R^2} g(y)d^2 y)\ln r \; +\; O(1/r)
\end{equation}
and we define
\begin{equation}
\beta_{1} :=2\int_{R^2} g(y)d^2 y \;\;\;\; \beta_1 \geq 0
\end{equation}
Substituting this into (38),
we get
\begin{equation}
\Delta \lambda_{0} + e^{-2\lambda_{0}} f_{1}(x)=0\;\;\;\;
                        f_{1}(x)= f(x) e^{-2\lambda_{1}}
\end{equation}
Finally, let $u=-\lambda_0$. Then
\begin{equation}
\Delta u + K(x) e^{2u} =0, \;\;\; K(x)=- f_{1}(x)
\end{equation}
This is the desired form of the eqaution.  Thus, we have put
\begin{eqnarray}
u  & =  & - (\lambda  + \int_{R^2}g(y) \ln |\vec{x}-\vec{y}| d^2 y )\\
K(x) & = & - f(x) \exp{(2\int_{R^2}g(y) \ln |\vec{x}-\vec{y}| d^2 y)}
\end{eqnarray}
{}From (22) , (49) and (50) and the asymptotic behaviour of $P^{a}_{b}$
and the
fact that the matter fields are of compact support, the asymptotic behaviours
 of $u$ and $K(x)$ are:
\begin{eqnarray}
u & \sim & {(\beta - \beta_{1})\over  2} \ln r \; + \; O(1/r^{\epsilon})\;\;\;
                                           \epsilon >0 \\
K(x) &  \sim  & 1/ r^{l_{1}- \beta_{1}}   {\rm to\; nontrivial\;
     leading\; order\; with\; }l_{1} \geq 4
\end{eqnarray}
 Hence, we are looking for solutions to (48) with asymptotic behaviour (51)
for $K\leq 0$, $K$ not vanishing identically, with asymptotic behaviour of
$K$ given in (52).

This is exactly the form of the equation for `` Prescribed Negative Curvature
in $R^2$'' examined in \cite{mcowen}.
We quote the content of the main theorem of this paper below:\\
\noindent{\bf Statement A:} If $-C r^{-l}\leq K\leq 0$ where $C>0,\; l>2$ and
$K(x_{0}) < 0 $ where $x_{0}$ denotes a point in $R^2$,  then there exist
$C^{2}$ solutions to (48) with
\begin{equation}
u = \alpha  \ln |\vec{x}-\vec{x_{0}}|   + u_{\infty} +O (r^{\gamma})
         \;\;\;\; {\rm as }\; r \rightarrow \infty
\end{equation}
with $u_{\infty} $ being a constant, for every
$\alpha\; \epsilon \; (0,{l-2 \over 2})$ and every
  $\gamma  > max (-1, 2-l+2\alpha)$.

Moreover, if $K\leq -C_{0} r^{-l}$ for $r> r_{o}$ (where $C> C_{0} > 0$)
and  $\alpha \geq {l-2\over 2}$, then  (48) admits no solution satisfying
\begin{equation}
u = \alpha \ln r + O(1) \;\;\;  r \rightarrow \infty \nonumber
\end{equation}

Further, from \cite{sattinger}, we have \\
\noindent{\bf Statement B:}If $K\leq 0$  on $R^2$
and $K\leq -{C_{0}\over r^2}$
for $r\geq r_{0}, (C_{0}, r_{0} >0)$, then (48) admits no solution.

We make the following remarks:\\
\noindent{\bf (1)} By integrating both sides of (48) and using Stokes
 theorem we
conclude that solutions to (48) satisfying (51)
must have $\beta > \beta_{1}$.
It follows that $\beta=0$ corresponds uniquely to the 2+1
vacuum case (all matter fields and momenta vanish).
It also follows that $\beta >0$. Note that $K=0$ slices were used in a
similar way to prove positivity of energy  by Henneaux in \cite{marc}.  \\
\noindent{\bf (2)}(a) Given that a solution to (48)
exists with a {\em prescribed}
value of $\beta$, we do not know anything about  uniqueness
of this solution since we are dealing with unbounded domains. (b)Moreover,
 we are not interested in prescribed values of $\beta$,
instead, we would like $\beta$ to be {\em determined} by the equation (48),
as
in
the $f=0$ case.\\
\noindent{\bf (3)}The parameter $\alpha$ in statement A, is to be identified
with ${\beta-\beta_{1} \over 2}$and $l$ in statements A and B, with
$l_{1}- \beta_{1}$. Then A states that for $l_{1} > \beta_{1} +2$,\\
\noindent (i)solutions to (48) exist for
every $\beta$
in the open interval $ (\beta_{1} , l_{1}-2)$.\\
\noindent (ii)if $\beta > l_{1} -2$ and the angular dependence
of $K$ does not attenuate the leading order term in such a way as to
violate the inequality on $K$ in A,
then no solutions exist with behaviour (54).

Further, B states that with a similar assumption
on angular dependence of $K$,
that no solutions exist to (48) for $l_{1} \leq \beta_{1} +2$.\\
\noindent{\bf (4)}Note that $l_{1}$ is determined, essentially, by the
``multipole moments'' of the source term $M_{i}(y)$. Hence a whole range of
energies seem to be possible for the same initial data! This is unphysical.
The resolution to this is that in (22) there is {\em no constant} piece.
Note that
in the proof of A in \cite{mcowen}, the value of $u_{\infty} $ is not
independent
of $\alpha$. As a logical consequence of remark (2)(b) we are led to make the
following conjecture:\\
\noindent{\bf Conjecture 1:}Among the pairs $(\alpha, u_{\infty})$ in A,
if a pair $(\alpha_{0}, 0)$ exists, then it is unique (i.e. $u_{\infty}=0$
singles out the particular value of $\alpha=\alpha_{0}$).

If this conjecture is true, then the equation (48) itself
determines the energy $\beta= 2\alpha_{0} +\beta_{1}$. (In this way, by using
physical intuition we could make more mathematical conjectures about the
equation).\\
\noindent{\bf (5)}Note that (48) has the following property: If (48) has a
solution $u_{1}(x)$ for $K=K_{1}(x)$ then $u_{2}(x) = u_{1}(x/c)$ is a
solution to
(48) with $K= {K_{1}(x/c)\over c^{2}}$.
We do not have a clear idea as to how to
use this property, but simply note that \\
$u_{2\infty}= u_{1\infty} - 2\alpha_{1} \ln c ,\;\;\;\; \alpha_{1}=\alpha_{2}$
 in obvious notation.\\
\noindent{\bf (6)}There is no reason to expect $\alpha_{0}$ in the
conjecture above to be such that $\beta <2$ and we must discard all
initial data which would
violate $\beta< 2$. However, note that in the ``generic'' case, $l_{1}=4$
and A says that $\beta<2$!

\section{Preservation of the gauge fixing under evolution}
Preserving the condition $P=0$, leads to the following equation for the lapse,
just as in \cite{vince}:
\begin{equation}
\Delta N = pN\;\;\;p:= e^{-2\lambda}[P^{a}_{b}P^{b}_{a}+ {s^{2}\over 8}
                                           +{e^{4\gamma}\over 2}v^{2}]
\end{equation}
Preservation of the conformal flatness condition is equivalent to demanding
that $\sqrt{h} h^{ab}$ be preserved and that $\lambda$ be determined by the
 Hamiltonian constraint at each instant of time (see discussion
in \cite{vince}).
This leads to the following equation for the shift:
\begin{equation}
\sqrt{h} (\partial_{a}N^{b} +h^{cb}h_{da}\partial_{c}N^{d}-\delta_{a}^{b}
                                \partial_{c} N^{c})= NP^{bc}h_{ca}
\end{equation}
Since transverse traceless tensors vanishing at spatial infinity , vanish
everwhere on $R^{2} $ using an argument similar to that of the lemma in
section 3, one can obtain an equation equivalent to the above equation by
taking its divergence to get:
\begin{equation}
\Delta N^{a} = -\partial_{b} (2N e^{-2\lambda}h^{bc} P^{a}_{c})
\end{equation}
In what follows, we assume the required smoothness on functions so that
we can apply results from standard elliptic theory.

\subsection{The equation for the lapse}
Near infinity $p\sim 1/r^{4-\beta}$. Since we are interested only in the case
$\beta < 2$,
\begin{equation}
p\sim 1/r^{2+\epsilon} \;\; \;\; \epsilon > 0
\end{equation}
where $\epsilon$ can be arbitrarily small.

To keep as much  of a spacetime interpretation as possible,
we would like $\;\;$
$N>0$. For the case $p=0$ the only strictly
positive solution which satisfies boundary conditions on the lapse from
section 2, is $N=a$ where $a$ is a positive constant whose value we
can fix to unity.

{}From now on, we only concentrate on the  case in which $p$ does not vanish
identically. Assuming $N>0$
 and noting that $p\geq 0$, we integrate both sides of the
lapse equation over $R^{2}$ to get
\begin{equation}
\oint_{r\rightarrow \infty}{\partial N\over \partial r} r d\theta=
      \int_{R^2}Npd^2 x \neq 0
\end{equation}
 Thus $N$ must diverge at least as
$\ln r$ as infinity is approached! But fortunately our framework allows such
lapses (pun not intended!). Since we expect the evolution to be generated
by the true Hamiltonian functional of the system, we look for solutions to the
lapse equation with asymptotic behaviour
\begin{equation}
N = a +b\ln r + O(1/r^{\epsilon}),  \;\;\;\epsilon, a,b>0
\end{equation}
Note that the condition $a>0$, comes from demanding that the ``non gauge''
part of the evolution be generated in the forward time direction
(this will be discussed more, later). Using the linearity of the
equation we can fix $a=1$. So we
look for solutions to the lapse equation with asymptotic behaviour
\begin{equation}
N = 1 +b\ln r + O(1/r^{\epsilon}),  \;\;\; b, \epsilon >0,
\end{equation}
We note the following:\\
\noindent{\bf (1)}If a solution exists with the required asymptotic behaviour,
 then by the maximum principle, since it is strictly positive asymptotically,
 it is strictly positive everywhere.\\
\noindent{\bf (2)}The solution to the lapse equation satisfying (61) is
unique if it exists. We show this in two steps\\
\noindent(i) Let there be 2 distinct solutions $N_{1}, N_{2}$ with the same
value of $b$. Then an application of the maximum principle to their
difference
implies uniqueness.\\
\noindent(ii)Let there be 2 solutions $N_{1}, N_{2}$ with $b_{1}\neq b_{2}$
in obvious notation. Then a linear combination of $N_1$ and $N_2$ exists
(call it $N_3$) which approaches a positive constant near infinity and which
is a solution to (55), due to the linearity of the equation. By the maximum
principle $N_{3}>0$ everywhere. But we have shown that if $N_3 >0$
everywhere then
$N_3$ must diverge at infinity. Hence $b_{1}=b_{2}$ and by (i) , $N_3 =0$
everywhere.\\
\noindent{\bf (3)}Using the method of sub and supersolutions we have made
partial progress on the question of existence of solutions to the lapse
equation. We
describe the details in the appendix. The result is that for weak enough
matter fields and momenta a solution does exist to (55) with required
asymptotic
behaviour. For arbitrary initial data, we have been able to show that a
solution exists
with asymptotic behaviour such that as $r\rightarrow \infty$
\begin{equation}
\delta\ln r - a_{1} < N < \delta \ln r +a_{2} , \;\;\;\;\delta, a_{1}, a_{2} >0
\end{equation}
where $\delta, a_{1}$ and $a_{2}$ have values given in the appendix.

\subsection{The shift equation}
Let
\begin{equation}
f^{ba}:=-2N e^{-2\lambda}h^{bc} P^{a}_{c} \;\;\;
\;f^{ri}=f^{ba}(dr)_{b}(dx)_{a}
\end{equation}
Note that
\begin{equation}
f^{ri}\sim 1/r^{\epsilon} \;\;\;\; \epsilon =(2 -\beta)>0
\end{equation}
Since $0\leq \beta <2 $, $\;\;0< \epsilon \leq 2$.

We first examine the case in which $\epsilon < 2$.
We define a solution $u_{R}^{i}(x)$ to  (57) for $(r= |\vec{x}|)<R$ in a
ball of radius
$R$ centred on the origin  and denoted by $B_{R}(0)$:
\begin{eqnarray}
2\pi u_{R}^{i}(x) & = &
    [  \int_{B_{R}(0)}f^{\bar{r} i}(\bar{x})d\bar{\theta} d\bar{r}
   -  \oint_{\bar{r}=R}f^{\bar{r} i}(\bar{x})
(\ln \bar{r})\bar{r}d\bar{\theta} ] \nonumber \\
                  &  & +\int_{B_{R}(0)} {\partial f^{ki}(\bar{x})
\over \partial {\bar{x}}^{k}}\ln |\vec{x}-\vec{\bar{x}}|
                        d^2 \bar{x}
\end{eqnarray}
A straightforward, but lengthy analysis shows:\\
\noindent(1) For fixed finite $r$
and every  finite $R$ such that $R> 1$ and $ R> (3/2)r$,
\begin{eqnarray*}
|u_{R}^{i}(x)| & < M(r,R)
\end{eqnarray*}
where $M$ is a finite number depending on $R$ and $r$ (we could try and
remove the restriction $R> (3/2)r$, but since it suffices to deal with
such $R$ in our subsequent arguments, we have not attempted to do so).\\
\noindent(2)For every $r$, there exists an $R_{0}$ such that for every
$R>R_{0}$,
\begin{eqnarray*}
|u_{R}^{i}(x)| & < M(r,R_{0})
\end{eqnarray*}
Moreover, for $R_{0} \rightarrow \infty$, $M(R_{0},r)\rightarrow f(r)$
such that $f(r)$ is finite for every finite $r$ and
as $r\rightarrow \infty$, the behaviour of $f(r)$ to leading order is
\begin{eqnarray}
f(r) & \sim & r^{1-\epsilon} + C_{\epsilon} \;\;{\rm for}\; \epsilon \neq 1 \\
     & \sim & (\ln r)^{2}  \;\; {\rm for}\; \epsilon=1
\end{eqnarray}
where $C_{\epsilon}$ is a finite  constant depending on $\epsilon$. Also,
$2>(\epsilon =2-\beta)>0$.

%Consider the compact sets $\bar{B}_{R_{n}}(0)$
%(i.e. the closed disks of radius $R_{n}$ centred at the origin) for
%$R_{n}= 2^{n}, \;\; n=0,1,2...$ and the sequence of solutions
%$u_{R_{m}}$
%in the nested open balls
%$B_{R_{m}},\;\;R_{m}= 2^{m}, \;\; m=1,2,3...$. Given a compact set
%$\bar{B}_{R_{n}}(0)$, the sequence of solutions
%$u_{R_{m}}$  for $m>n$ restricted to
%$\bar{B}_{R_{n}}(0)$ is uniformly bounded by (1) and (2) above.
Using standard
arguments which involve the construction of
appropriate sequences of solutions from the above 1 parameter set and which
invoke the Arzela-Ascoli theorem (for this theorem, see, for example
\cite{ascoli}), we are guaranteed the existence of a solution to (57)
with asymptotic behaviour, at worst, as in (66) and (67).
Note
that the norm of the shift evaluated with the metric $q_{ab}$ goes to
zero near infinity.

 Uniqueness of this solution upto addition of a constant, $c^{i}$ follows
from the fact that the Laplace equation in 2 dimensions
admits no solutions with sublinear growth near infinity except the constant
solutions.

For the case, $\epsilon =2$, the only possible data is that for vacuum 2+1
gravity (see remark 1, section 5.2). In this case $f^{ab}=0$ and the shift
equation admits only constant solutions.

The nonuniqueness upto addition of a constant is also present in the
compact case \cite{vince}. Here, it comes about because we have a freedom
to specify the coordinate system on each slice upto a translational
isometry of $h_{ab}$ (note
that we do not have a similar `rotational' freedom because rotations are
generated by the conserved angular momentum (see \cite{asht}) and {\em not}
by the
diffeomorphism constraints). We can fix this nonuniqueness, as in \cite{vince}
by  imposing  an appropriate condition on the shift, say that the `constant'
part of the shift  near infinity vanish.

\subsection{Discussion}
We are forced to use lapses and shifts which diverge at spatial infinity.
It is
questionable whether we accept such behaviour from a spacetime viewpoint.

In 3+1 dimensions, spacetime intuition and the constraint theory
interpretations agree nicely. Evolutions which do not move the
spatial manifold  with respect to the fixed structure at infinity are
interpreted as gauge from the constraint systems point of view and this
is natural from a spacetime viewpoint. In the
2+1 dimensional case studied here, we have divergent motions in space-time,
of
the points of the spatial manifold near spatial infinity. Yet, these
motions are interpreted as gauge (if $a=0$ in (60)) from a constraint
theory viewpoint.
We do not understand, in any deep way, how this comes about.  But we emphasize
that {\em viewed purely as a constrained dynamical system, the formalism
developed is self consistent}. We do admit, however, that we do not know if a
rigorous treatment of the evolution equations
using appropriate function spaces (such as in \cite{christo} for the 3+1 case)
would show an inconsistency in the framework developed in section 2.

With regard to the permissible behaviour of the lapse, there may be an even
more dramatic clash of spacetime interpretation and constraint theory
viewpoint, than that alluded to above. If the lapse equation (55) admits
solutions with asymptotic behaviour as in (60) {\em but with $a<0,\; b>0$},
the ``non gauge'' part of the evolution is backward in time.
But since $b>0$, the ``gauge'' part of the evolution
dominates this completely and the 2-slices are still
pushed forward in time!

We beleive that the conformal flatness gauge fixing is
the most natural one to impose since,
not only is the 2 manifold $R^{2}$, but in addition, this
gauge fixing interacts well with the boundary conditions on the 2-metric,
which are
also conformally flat (to leading order). The maximal slicing condition
simplifies the solving of the constraints and ensures that $K(x)\leq 0$
in (48),
so that we can use the results of
\cite{mcowen,sattinger}. It also simplifies
part of the problems one faces in the attempt to define a reduced
Hamiltonian description (see section 7). Unfortunately, the
corresponding spacetime picture is
not very appealing, although there seems to be something to be
understood here.

\section{The reduced Hamiltonian description}

To go to a reduced Hamiltonian description, we will {\em assume} that:\\
\noindent(i) the Hamiltonian constraint can be solved uniquely for $\lambda$
with $\beta$ determined by (48).\\
\noindent(ii) the lapse equation (55) has a solution with the asymptotic
behaviour in (61).\\
If (i) and (ii) are true, we can eliminate the constraints by expressing the
gravitational variables in terms of the matter variables (the latter
parameterize the true degrees of freedom of the theory). Using the
gauge fixing
conditions and eliminating the constraints from the action (15),
we obtain the
reduced action:
\begin{equation}
S_{red}= \int dt [(\int_{R^2}d^2 x \;\; s\dot{\gamma}\;+\; v\dot{\omega})
                         \;\; -\;\; 2\pi\beta]
\end{equation}
where $\beta$ is to be understood as a functional of the matter fields and
momenta. The reduced Hamiltonian is then,
\begin{equation}
H_{red}\;\; = \;\;2\pi \beta[\gamma, \omega, s, v]
\end{equation}
Note that unlike in \cite{vince}, this is a time independent Hamiltonian.
Thus, given initial data satisfying the integrability condition (34), we have
reduced the system (assuming (i) and (ii) above) to that of two matter fields
whose time evolution is determined by a non-explicit Hamiltonian. (One of the
reasons we worked on $R^2$ is that we could {\em explicitly} solve
the diffeomorphism constraints.)

This suggests
that if we could set up some sort of formal perturbation scheme to solve (38),
we could try to use machinery from perturbative quantum field theory to
examine the quantum theory. One usually applies perturbative quantum field
theory to systems in which
one cannot solve the field equations nonperturbatively, but for which the
Hamiltonian is known explicitly in terms of the fields. Here one has the
additional complication that the Hamiltonian is not explicit but (maybe)
can also be written as a formal perturbation series. So given the
perturbation  expansion of the Hamiltonian in terms of the
matter fields, one would expect a more complicated combinatorics in
calculating transition amplitudes.
The perturbation expansion would have to be
supplemented with the condition that the total energy be bounded.
How one might do this is an interesting question whose
answer may lead to a sensible perturbative quantum field theory.
(Note that (34) would also have to be imposed). Admittedly, these remarks are
of a vague and speculative character.

In any event, the work in this paper has lead to a form of
the classical theory such that any attempt at quantization must face,
head on, the issue of upper
boundedness of the energy.

\section{Conclusions and open issues}

We have worked towards a reduced Hamiltonian description of 1 spacelike
Killing vector field reductions of 4 dimensional vacuum general relativity.
The reduced Hamiltonian is not explicit, but determined by the solution to
(48). This equation has been studied by mathematicians in the guise of the
equation
for prescribed negative curvature in $R^2$. There is presumably more in the
mathematics literature than \cite{mcowen,sattinger}, which could lead to
a better understanding of (48). The physical interpretation of (48)
led us to a mathematical conjecture  (see section 5) which (hopefully) can be
 validated in the future.

There is also an open question with regard to existence of solutions to the
lapse equation. Since this equation is linear, it is hoped that the question
can be answered by workers more mathematically knowledgeable than this author.

Balancing the good news that some
properties of (48) are known to mathematicians, is the bad news
regarding the spacetime interpretation of this work. It may be that one can
choose better gauge conditions (although
it has been argued in section 6 that the ones chosen in this paper seem to
be the simplest) so that no divergent lapses or shifts are encountered. One
possibility is to turn for inspiration to the cylindrical
waves analysed in \cite{karel}.

The cylindrical wave spacetimes may be viewed as 1 (z-directional,
translational) kvf reductions with an additional rotational kvf
(see discussion in \cite{asht}). Thus the system is equivalent to
rotationally  symmetric 2+1 gravity coupled to a single
rotationally symmetric scalar field
(in this midisuperpace the twist of the
translational kvf vanishes and the single scalar
field is related to the norm of the translational kvf).
The gauge conditions used in \cite{karel} are not
the same as those used in this work.
Our gauge conditions particularized to the
rotationally symmetric case lead to a Hamiltonian constraint
which seems to be difficult to solve in closed form. Thus our gauge conditions
are not adapted to the rotational symmetry, in contrast to the ones used in
\cite{karel}. In fact the
ones used in \cite{karel}
permit an exact and complete closed form solution to the equations.

Note, however, that there is an additional subtlety when one compares our
work with that in \cite{karel}. Because the asymptotic conditions we use are
{\em different} from those in \cite{karel}, so are the permitted variations
of the various fields in phase space. Thus we use  a (subtly) different
symplectic structure as compared to \cite{karel}, the difference coming from
what we can and cannot hold fixed at infinity.
It would clarify our gauge fixing
conditions if maximal slices ( in the 2-d sense) could be explicitly
constructed in the cylindrical wave spactimes.

One could also try and generalize the cylindrical wave boundary conditions
and
gauge fixings to the general 1 (translational) kvf case. We plan to look at
this issue in the future.

Finally, we restricted our considerations to $\Sigma = R^2$ mainly for
simplicity and so that we could explicitly solve the diffeomorphism
constraints.
We hope that some progress can be made towards a quantum theory along the
(extremely vague) lines sketched out in section 7.

\begin{center}
{\Large\bf  Appendix}
\end{center}

\appendix
\subsection*{A. Subsolution to the lapse equation}
Given $B>1$,
\begin{equation}
p(x)\leq {A \over (B+r^{\delta})^{2 +{2\over\delta}}}
\end{equation}
for suitable $A,\delta >0$ and $\delta <1$. Consider the two cases:\\
\noindent{\bf (i) ${2A\over B\delta^{2}}\leq 1$:} This corresponds
to ``small''
initial data. In this case a subsolution is
\begin{equation}
N_{sub}= \ln (B+ r^{\delta})\;\;+\;\;1-e^{-1}
\end{equation}
where $e$ is Euler's constant. Note that
the maximum value of $\ln r / r$ for $r>0$
is $1-e^{-1}$. Using this it is straightforward to check that
\begin{equation}
\Delta N_{sub} > q(x)N_{sub}
\end{equation}
as required with
\begin{equation}
N_{sub} \sim \delta \ln r\; +\; 1-e^{-1}
\end{equation}
as infinity is approached.\\
\noindent{\bf (ii) ${2A\over B\delta^{2}}> 1$:} This corresponds to the
generic case. We choose
\begin{eqnarray}
N_{sub} & =  & \ln (B+ r^{\delta})- \ln (B+ R^{\delta})\;\;{\rm for}\; r>R\\
        & =  & 0    \;\; {\rm for}\; r\leq R
\end{eqnarray}
where $R$ is large enough that for $r>R$,
\begin{equation}
\Delta N_{sub} = {B\delta^{2} r^{\delta-2} \over 2(B+r^{\delta})^{2}}
                    \geq {A \over (B+r^{\delta})^{2 +{2\over\delta}}}
                              \ln (B+ r^{\delta})
\end{equation}
Thus asymptotically
\begin{equation}
N_{sub} \sim \delta \ln r\; - \; \ln (B+R^{\delta})
\end{equation}

\subsection*{B. Supersolution to the lapse equation}
We concentrate only on the $p\neq 0$ case. Then there exists a ball
of radius $\epsilon$ around a point $x_{0}$ denoted by $B_{\epsilon}(x_{0})$
such that for $x$ in $B_{\epsilon}(x_{0})$, $p(x)\geq k>0$. Define
$\phi(x)$ to be a smooth function with support on $B_{\epsilon}(x_{0})$,
such that $0\leq \phi <(k/2\pi )$ and $\phi$ does not vanish identically.
Define
\begin{eqnarray}
y_{0}(x) & = & \int_{B_{\epsilon}(x_{0})}
\phi(\bar{x})\ln |\vec{x}-\vec{\bar{x}}|
                        d^2 \bar{x} \\
y_{1}(x) & = & y_{0}(x)- inf[B_{\epsilon +1}(x_{0})] +2
\end{eqnarray}
Then $\Delta y_{1}(x)\leq q(x)y_{1}(x)$ and $y_{1}(x)$ is a supersolution
with asymptotic behaviour
\begin{equation}
y_{1}(x)\sim \int_{B_{\epsilon}(x_{0})}\phi(\bar{x})d^2 \bar{x}
                            \ln r + 2 -
inf[B_{\epsilon +1}(x_{0})]
\end{equation}
where $inf[B_{\epsilon +1}(x_{0})] $ refers to the infimum of $y_{0}(x)$
in the ball of radius $(\epsilon +1)$ centred at the point $x_{0}$.
If $y_{1}(x)$ is a supersolution to the linear lapse equation, then so is
$y_{2}= {y_{1}\over d} +e,\;\;d,e>0$. Hence, for case (i) we can choose
$d,e$ in such a way that $N_{sup}$ has asymptotic behaviour
\begin{equation}
N_{sup} \sim \delta \ln r +a_{2}\;\;\; a_{2}>1-e^{-1}
\end{equation}
By standard arguments in the method of sub and super solution using the
maximum principle and the Ascoli-Arzela theorem \cite{ascoli}, we are
guaranteed
existence of a solution, $N$, to the lapse equation with asymptotic
behaviour such that
\begin{equation}
\delta \ln r + (1-e^{-1})\leq N \leq  \delta \ln r + a_{2}
\end{equation}
For case (ii), using the same standard arguments,
we can  show existence of a solution  with asymptotic
behaviour such that
\begin{equation}
\delta \ln r - \ln(B+R^{\delta})\leq N \leq  \delta \ln r + a_{2}
\end{equation}

{\bf Acknowledgements:} I am indebted to Andrejs Treibergs
for discussions related to elliptic pdes and for showing me the reference
\cite{mcowen}.
I would like to thank Karel Kucha\v{r}, Joe Romano and Richard Price for
valuable discussions. This work was supported by the NSF grant PHY9207225.

\end{document}